\documentstyle{mn}
\begin{document}
\def\lsun{{L_{\odot}}}
\def\msun{{\rm M_{\odot}}}
\def\rsun{{R_{\odot}}}
\def\be{\begin{equation}}
\def\ee{\end{equation}}
\def\m2i{M_{2,\rm i}}
\newcommand{\lta}{\la}
\newcommand{\gta}{\ga}
\input{epsf.sty}
\def\plotone#1{\centering \leavevmode
\epsfxsize=\columnwidth \epsfbox{#1}}
\def\plottwo#1#2{\centering \leavevmode
\epsfxsize=.45\columnwidth \epsfbox{#1} \hfil
\epsfxsize=.45\columnwidth \epsfbox{#2}}
\def\plotfiddle#1#2#3#4#5#6#7{\centering \leavevmode
\vbox to#2{\rule{0pt}{#2}}
\includegraphics{#1}}
\def\spose#1{\hbox to 0pt{#1\hss}}
\def\lta{\la}
\def\gta{\ga}
\def\Msun{{\rm M}_\odot}
\def\msun{{\rm M}_\odot}
\def\rsun{{\rm R}_\odot}
\def\Lsun{{\rm L}_\odot}
\def\half{{1\over2}}
\def\RL{R_{\rm L}}
\def\zs{\zeta_{s}}
\def\zR{\zeta_{\rm R}}
\def\dJJ{{\dot J\over J}}
\def\dMM{{\dot M_2\over M_2}}
\def\tKH{t_{\rm KH}}
\def\eck#1{\left\lbrack #1 \right\rbrack}
\def\rund#1{\left( #1 \right)}
\def\wave#1{\left\lbrace #1 \right\rbrace}
\def\dd{{\rm d}}
\def\be{\begin{equation}}
\def\ee{\end{equation}}
\def\mt{M_{\rm tr}}
\def\lb{\label}
\def\d{{\rm d}}
\def\ep{\epsilon}

\title{The evolution of black--hole mass and angular momentum 
}
\author[A.~R.~King \& U.~Kolb]{A.R.~King and U.~Kolb\\ 
Astronomy Group, University of Leicester, Leicester LE1 7RH, U.K. \\
(ark@star.le.ac.uk, uck@star.le.ac.uk)}


\maketitle

\begin{abstract}
We show that neither accretion nor angular momentum extraction are
likely to lead to significant changes in the mass $M_1$ or angular momentum
parameter $a_*$ of a black hole in a binary system with realistic
parameters. Current values of $M_1$ and $a_*$ therefore probably
reflect those at formation.
We show further that sufficiently energetic jet ejection powered by
the black hole's rotational energy can stabilize mass transfer in
systems with large adverse mass ratios, and even reduce the mass
transfer rate to the point where the binary becomes transient. 
\end{abstract}


\section{INTRODUCTION}
\label{sec:intro}

In a recent paper, Zhang et al.\ (1997)  
use the observed strength of the ultrasoft X--ray 
component in black--hole binaries to estimate the black hole
spin. They argue that known systems show a range of spin 
rate $a_* = a/M_1$ (see their Table 2), where 
the Kerr parameter $a = cJ_1/GM_1$ with $M_1,J_1$ the black--hole mass
and angular momentum. In particular the two superluminal jet
sources GRO~J1655-40 ($a_* \simeq 0.93$)
and GRS 1915+105 ($a_* \simeq 0.998$) are claimed to spin at
rates close to the maximum value $a_*=1$, while systems such as
the soft X--ray transient GS 2000+251 have $a_* \simeq 0$. 

The obvious question is whether this claimed range of spin rates
reflects systematic spinup from $a_*=0$, or
spindown from $a_*=1$, or the accidents of birth.
In this paper we shall show that neither spinup nor spindown is likely
to account for the range. The 
total mass accreted over the lifetime of any binary black hole
is too small to increase $a_*$ from 0 to 1. By
contrast, spindown of a black hole from an initial state with $a_* =1$
is a relatively efficient process
if the rotational energy is used to power moderately
relativistic jets. However, significant spindown requires the
hole to have released improbably large amounts of energy into the
surrounding interstellar medium. 

It seems therefore that binary black
holes retain rather similar spin rates to those they had at
birth. The same appears true of their masses: although a
loss of rotational energy from a black hole
implies that its gravitating mass decreases, we shall see that
with realistic energy extraction efficiencies this effect is rather
small. We conclude that binary black holes essentially retain the masses and
spin rates they were born with. A further interesting result emerges
from our analysis: moderately relativistic jet ejection is able to
stabilize mass transfer in binaries where the mass ratio would
otherwise lead to unstable Roche lobe overflow. Sufficiently energetic
jet ejection can even lower the mass transfer rate to the point where
the system becomes transient.

\section{TOTAL MASS TRANSFER IN BLACK--HOLE BINARIES}
\label{sec:mt}

A black hole can change its spin either by accreting mass and
the associated angular momentum, or by giving angular momentum
to matter in its close vicinity. If the hole is a member of a
binary system both of these processes are
limited by the total mass transferred in the system's lifetime.
In this section we estimate this quantity for all types of
black--hole binary. We can divide these into high--mass 
systems, where mass is accreted from the stellar wind of an
early-type companion, and low--mass systems, where a late--type
star fills its Roche lobe. We may further subdivide the latter
group into those where the mass transfer is driven by the 
nuclear or thermal expansion of the secondary (n--driven), and those
where the driving mechanism is angular momentum loss (j--driven). 

\subsection{Wind--fed binaries}
In this type of binary the black hole generally accretes 
only a small fraction $f=\dot M_{\rm acc}/\dot M_{\rm w}$ 
of the total mass lost by the companion, usually a
supergiant, in the form of a wind at a rate $\dot M_{\rm w}$.
A rough estimate for $f$ is
\be
 f \simeq \frac{r_a^2}{4 a^2} \simeq \frac{G M_2^2}{a^2 v_{\rm w}^4}
 \simeq 2 \times 10^{-3} {\left( \frac{m_2}{P_d} \right)}^{4/3} 
 {\left( \frac{v_{\rm w}}{1000\, {\rm km}\, {\rm s}^{-1}}\right)}^{-4}
\label{0}
\ee
(e.g.\ Bhattacharya \& van den Heuvel 1991), where $m_2 = M_2/\msun$
is the companion mass in solar masses, $a$ the orbital separation, $P_d$
the orbital period (in days), $r_a$ the accretion radius and $G$
the constant of gravitation. 
With wind velocities $v_{\rm w} \la 1000$~km s$^{-1}$ (e.g.\
Behnensohn et al.\ 1997) we find typically $f \la 10^{-2}$.


Even with extreme assumptions it is clear that the total transferred
mass $\mt$ cannot exceed a quantity of order
\be
\mt({\rm wind}) \sim 0.1\msun
\label{1}
\ee

\subsection{Roche--lobe overflow systems}

In such systems a significant fraction of the secondary mass $M_2$ can be
transferred, i.e.
\be
\mt({\rm Roche}) \lta M_2,
\label{1a}
\ee
subject only to the requirement that mass transfer should remain stable
(i.e. does not become so rapid that the system enters a
common--envelope phase). This mass is transferred on a timescale 
$t_{\rm ev}$, where $t_{\rm ev}$ is the evolution time of the 
binary in the mass--transferring stage. This is given by 
the angular momentum loss time $t_{\rm j}$ for j--driven systems
(donor a main-sequence star and $M_2\la1.5\msun$), and 
either the thermal timescale for donors crossing the Hertzsprung gap  
(Kolb et al., 1997; Kolb, 1998) or the nuclear lifetime $t_{\rm n}$ 
if the donor is a low--mass giant or a main--sequence star with mass
$\ga 1.5\msun$.
Not all of the transferred mass
may be accreted by the black hole however, as the
accretion rate cannot consistently exceed the Eddington value 
$\dot M_{\rm Edd} = 10^{-8}m_1$ $\msun$ yr$^{-1}$, where 
$m_1 = M/\msun$. An upper limit to the {\it accreted} mass is 
therefore in all cases given by
\be
M_{\rm acc}({\rm Roche}) = {\rm min}[M_2,\  10^{-8}m_1t_{\rm ev}\msun]
\label{2}
\ee

In practice this means $M_{\rm acc} \la \mt < M_2 < 1.5\msun$ for
j--driven 
(short--period) systems. For n--driven systems we plot an estimated
upper limit for $M_{\rm acc}$ as a function of initial donor mass in
Fig.~1 (case A mass transfer, donor on the main sequence) and Fig.~2 
(case B mass transfer, post--core hydrogen but pre--core helium
burning phase). 
The initial black hole mass is $8\msun$ in all cases; systemic angular
momentum losses are assumed to be negligibly small.
We used simple fitting formulae to describe the variation
of global stellar parameters along single--star tracks as given by Tout
et al.\ (1997). To calculate the case A mass transfer rate the
donor's radius expansion rate $K={\rm d}\ln R/{\rm d}t$ 
was approximated by that of a single star with the donor's age and 
current mass. For case B 
the core mass growth determines the radius variation with little
sensitivity to the total mass (e.g.\ Webbink et al.\ 1983 for 
low--mass stars; Kolb 1998 for stars of higher mass), hence we used
$K(t)$ from a single star with the donor's initial mass to estimate
the transfer rate.
The accreted mass $M_{\rm acc}$ shown in Figs.~1 and 2 represents an
upper limit as mass transfer could begin at a later phase than assumed
(donor on the ZAMS for case A, on the terminal main sequence for case
B). A system formally terminating case A mass transfer when the donor
arrives at the terminal main sequence would continue to transfer mass
via case B. The case B phase terminates when core helium burning
begins or the donor's envelope is fully lost.  
Any mass transferred in excess of the Eddington rate was
assumed to leave the system with the black hole's specific orbital
angular momentum, otherwise mass transfer was taken to be
conservative.

\section{CHANGING THE BLACK--HOLE SPIN BY ACCRETION}

Here we consider the effect of accretion in changing the black hole
mass and spin.
We assume first that black--hole spinup occurs by accretion 
from a disc terminating at the last stable circular orbit.
The accreting matter adds both its rest--mass and its 
rotational energy to the hole, increasing both the gravitating
mass--energy $M_1$ of the hole and its angular momentum $J_1$. 
Bardeen (1970; see also Thorne, 1974) showed that these 
quantities increase as
\be
\Delta M = 3M_i[\sin^{-1}(M_1/3M_i) - \sin^{-1}(1/3)],
\label{3}
\ee
\be
a_* = \biggl({2\over 3}\biggr)^{1/2}{M_i\over M_1}
  \biggl[4 - \biggl({18M_i^2\over M_1^2} - 2\biggr)^{1/2}\biggr].
\label{4}
\ee
Here $\Delta M$ is the rest--mass added to the hole from
the initial state (assumed to be $M_1=M_i, a_*=0$). Once this
is such that $M_1/M_i = 6^{1/2}$ we see from (\ref{4}) that 
$a_*=1$. In practice the value of $a_*$ stops slightly short
of this maximal value, and further accretion simply maintains
this state (Thorne, 1974). From (\ref{3}) we see that the
required additional rest--mass is 
\begin{eqnarray}
\Delta M & = & 3[\sin^{-1}(2/3)^{1/2} - \sin^{-1}(1/3)]M_i \nonumber
 \\ 
 & \simeq & 1.85 M_i = 0.75 M_1 \label{5}
\end{eqnarray}
Figure~3 shows $a_*$ as a function of $\Delta M$ (given by numerically  
combining eqs. \ref{3}, \ref{4}). 
In order to spin up from  $a_*=0$ to a value $\simeq
1$ the black hole must have accreted a rest--mass of order
75\% of its current gravitating mass. This is therefore a 
lower limit to the accreted mass $M_{\rm acc}$.

The resulting lower limits to $M_{\rm acc}$ 
for the two presumed rapidly--spinning
systems GRS 1915+105 and GRO~J1655-40 can be compared with the
upper limits obtained from considerations as in Section 2 (see also
Tab.~1):

The binary period and component masses of GRS 1915+105 are not
known. Zhang et al.\ (1997) and Cui et al.\ (1998) find that 
consistency between black hole parameters derived from the X--ray
spectrum and from the $67$~Hz QPO (Morgan, Remillard \& Greiner 1997),
when interpreted either as a trapped g--mode oscillation or 
a disc precession from the frame--dragging effect, implies a high--mass
near--Kerr black hole ($M_1\simeq 30\msun$, $a_* \simeq 0.998$).
If this is true the lower limit on $\Delta M \ga
22\msun$ required for spinup is much larger than what could have
been transferred even in the most favourable evolutionary
configuration. Repeating the calculations shown in Figs.~1 and 2 for a
$30\msun$ primary gives $M_{\rm acc} \la 4\msun$ (case A) and $\la
1.5 \msun$ (case B).

Conversely, in GRO~J1655-40 the binary parameters are reasonably well
determined (Orosz \& Bailyn, 1997; van der Hooft et al., 1998;
Phillips et al.\ 1998). 
To find an upper limit to $M_{\rm acc}$ we assume 
that the past evolution was conservative, i.e.\ that $M_1^3M_2^3P
=$~const. Then the present system parameters as given by Orosz \&
Bailyn (1997), $M_1\simeq7\msun$, $M_2\simeq2.3\msun$, $P=2.62$~d,
imply a minimum period $P\simeq 1.08$~d. This is somewhere 
in the middle of the main sequence band (see e.g.\ Fig.~1) where the
donor's mass--radius index $\zeta$  
relevant for stability against thermal--timescale mass
transfer is $\zeta \simeq 0$ (Hjellming 1989). In this case mass
transfer stability demands that initially  $(M_2/M_1)_i<5/6$,
see (\ref{15}) below, hence initially $M_2 \la 4.2 \msun$, and therefore
$M_{\rm acc} \la 1.9 \msun \simeq 0.27\, M_1$.
(If $M_2$ were larger thermal--timescale mass transfer would ensue,
at a rate $\simeq M_2/t_{\rm KH} = f_{\rm Edd} \dot
M_{\rm Edd}$ with $f_{\rm Edd} = 10-100$; $t_{\rm KH}$ is the
donor's Kelvin-Helmholtz time. The black hole would accrete only a 
very small fraction $\simeq 1/f_{\rm Edd}$ of any transferred mass in
this phase). 
Using Phillips et al.'s (1998) lower limits for the present
component masses in GRO~J1655-40 ($M_1=4.2\msun$, $M_2=1.4\msun$) gives
a minimum orbital period of 1.11~d and also $M_{\rm acc} \la
1.9 \msun \simeq 0.27\, M_1$. This value is clearly inconsistent with the
spinup requirement $\Delta M > 0.47\, M_1 = 3.3 \msun$ 
if $a_*\simeq 0.93$, the value preferred by Zhang et al.\ (1997), and
only barely consistent with their lower limit $a_*> 0.7$ which
requires $\Delta M \gta 0.27\, M_1 \simeq 1.9\msun$. 

We conclude that the claimed range of
$a_*$ cannot be achieved by spinup of the black hole from an
initially non--rotating state.

Much attention has recently been paid to suggestions that quiescent
soft X--ray transient (SXT) systems might have higher accretion
rates than previously thought, because a large mass flux might
be advected into them, i.e.\ accreted at low radiation
efficiency. Advective flows have lower specific angular momentum 
than the Kepler value and so are even less effective in spinning up the
black hole. Advection therefore does not change the conclusion of the
last paragraph concerning spinup. However, since the advected specific
angular momentum is so low, one might consider the opposite
possibility, i.e. reducing $a_*$ to a value close to zero by diluting
the original angular momentum. 
However, even if the advected matter 
has zero angular momentum we have $a_* = J_{\rm i}/M_1^2 = 
a_*({\rm i})(M_{\rm i}/M_1)^2$, where $M_i$, $J_i$, $a_*(i)$ specify
the hole's initial mass, angular momentum and Kerr parameter,
respectively.  
Reducing $a_*$ from 1 to $\simeq 0.1$ in this way requires the black
hole mass to increase by a factor $\simeq 3$. Thus the transferred
mass must satisfy $M_{\rm tr} \gta 2M_1/3$, again far too large
compared with the limits (\ref{1}, \ref{2}).

\section{CHANGING THE BLACK--HOLE SPIN BY EJECTION}

A rotating black hole can lose angular momentum and rotational energy
because of the existence of an ergosphere outside its
event horizon. The energy loss implies that the gravitating mass of the
hole must decrease, according to
\be
\d M_1 = \ep \Omega_{\rm H}\d J
\lb{6a}
\ee
For convenience in this Section we use geometrized units, in which $G=c=1$.
Here 
\be
\Omega_{\rm H}  = {a_*\over 2M_1[1+(1-a_*^2)^{1/2}]}
\lb{7a}
\ee
is the apparent angular velocity of the horizon (e.g. Misner et al,
1973), and $\ep$ measures the efficiency of energy extraction. $\ep=1$
is the maximum value, and corresponds to a `reversible'
transformation, in which the area of the event horizon is held fixed,
while efficiencies $\ep \lta 0.5$ are typical for astrophysically
realistic processes (e.g. Blandford \& Znajek, 1977). Since
\be
\d J = \d (a_*M_1^2) = M_1^2\d a_* + 2a_*M_1\d M_1
\lb{8a}
\ee
we get with (\ref{6a}, \ref{7a}) 
\be
{2\d M_1\over M_1} = {\ep a_*\d a_*\over 1 - \ep a_*^2 + (1-a_*^2)^{1/2}}.
\lb{9a}
\ee
This can be integrated using the substitution $v = (1-a_*^2)^{1/2}$,
and gives finally
\be
M_1^2 = M^2_{\rm max}[(1-a_*^2)^{1/2} + 1]^{n}
\biggl[{\ep (1-a_*^2)^{1/2}  - \ep + 1\over 1 -\ep}
\biggr]^{m}
\lb{10a}
\ee
where 
\be 
n = -{\ep\over 2\ep - 1},
\lb{11a}
\ee
\be
m = -{\ep-1\over 2\ep - 1}
\lb{12a}
\ee
and the hole starts with $M_1 = M_{\rm max}, a_* = 1$. The maximum
rotational energy extraction is given by setting $a_*=0$ in
(\ref{10a}), which gives
\be
M_1^2 = 2^{n}(1-\ep)^{-m}M^2_{\rm max}.
\lb{13a}
\ee
Taking the limit $\ep \rightarrow 1$ gives $M_1^2 = M^2_{\rm max}/2$,
i.e. $M_1$ reaches its `irreducible' value $M_{\rm max}/\sqrt{2}$
(Christodoulou 1970, Christodoulou \& Ruffini 1971). This
maximal energy yield implies a reversible transformation in which the 
event horizon area is held constant. However in general the energy yield is
considerably smaller than this limit. Fig.~4 shows the extracted
fractional energy $\Delta m = (M_{\rm max} - M_1)/M_{\rm max}$ as a
function of $\ep$. We
see that for typical efficiencies $\ep \lta 0.5$ rather less than
about 10\% of the 
initial 
rotational energy is extracted; the hole's mass thus
remains very close to its original value $M_{\rm max}$. Physically
what is happening is that all of the hole's {\it angular momentum} is
extracted, but the associated {\it energy} is used inefficiently: the
extraction process allows much of this to disappear down the hole,
reducing the loss of gravitating mass.

The remaining energy must appear in some form outside the hole. 
We note that both the systems of Table 1
with large claimed $a_*$ are ejecting relativistic jets. These remove energy
at the rate $\Gamma \dot M_{\rm ej}c^2$, where
$\Gamma$ is the specific energy of the jet matter, and 
$\dot M_{\rm ej}$ is their mass--loss rate. If all the energy
of the jet material is in its bulk motion, $\Gamma$ is simply the
Lorentz factor $\gamma$ of this motion. However, if the spin energy of
the black hole is used in other ways, e.g. to excite relativistic electrons,
$\Gamma$ will exceed $\gamma$. Thus spindown wins out over 
spinup as a way of altering $a_*$ because the effect of transferring
rest mass $\mt$ from the companion is enhanced by a factor $\Gamma$ in
the former case. If all the 
transferred matter is ejected in this way until the hole is 
spun down, the requirement on the total transferred mass
becomes
\be
\mt = {\Delta m \over \Gamma}M_1.
\label{10}
\ee
Taking $\Delta m \sim 0.1, M_1 = 10\msun$ and bounding $\Gamma$ below 
with the value $\gamma \sim 2.55$ inferred for GRO~J1655-40
(Hjellming \& Rupen 1995) we get a limit $\mt \lta
0.4\msun$ on the mass which must be transferred to reduce the spin to zero. 
Spindown from $a_*=1$ to $a_*=0$ therefore seems possible for systems
with Roche--lobe overflow. However, if this has occurred in a given
system, one would expect to see
abundant evidence of the effects of the total extracted energy 
$\Delta m M_1 c^2 \simeq 10^{54}$ erg 
on the surrounding interstellar
medium. GRO~J1655-40 would deposit this energy over a time of $\simeq
10^6$~yr, see (\ref{18}) below. 
The mean energy output rate $\simeq 8 \times 10^6 \lsun$ is 
comparable to the luminosity of a cluster of 10 O/B supergiants.
The detectability of such an energy deposition depends on the
structure and density of the local interstellar medium. 
The fact that none of the claimed $a_* \simeq 0$ systems shows
such evidence suggests that jets are produced only over a relatively
short fraction of the system's lifetime. 
In this case $a_*$ is likely
to remain close to its original value ($\simeq 1$).

\section{BINARY EVOLUTION WITH RELATIVISTIC JETS}

We have seen above that the production of relativistic jets by a
rotating black hole can lead to a large loss of gravitating mass from
this object, i.e.\ $\dot M_1 ({\rm jet})= - \Gamma \dot M_{\rm ej}$,
where $\dot M_{\rm ej}$ is the rest mass the jet carries away per unit
time. As the jet ejection is a consequence of mass accretion, the
ejected rest--mass must be roughly equal to the transferred rest--mass
from the companion. 
Hence we expect that the ejection parameter $\eta$, defined by
$-\dot M_{\rm ej} = \eta \dot M_2<0$, is of order unity.
(Based on energy considerations for the observed jet
ejection, Gliozzi et al.\ 1998 argue that $\eta$ is close to 1 in
GRS~1915+105).
In addition the jets will presumably carry off the
specific orbital angular momentum $j_1 = M_2J/M_1M$ of the black hole
from the binary orbit, where $M, J$ are the total binary mass and
orbital angular momentum. The jet ejection process therefore
constitutes a `consequential angular momentum loss' or CAML
process. The effects of CAML on Roche--lobe--filling systems 
were studied quite generally by King \& Kolb (1995; henceforth KK95),
who specified them in terms of mass loss and angular momentum loss
parameters $\alpha, \nu$, with 
\be
\dot M_1 = (\alpha - 1)\dot M_2,\ {\dot J_{\rm CAML}\over J} =
\nu{\dot M_2\over M_2}.
\label{11}
\ee
Thus here we have 
\be
\alpha = \eta \Gamma,\ \nu = {\eta \Gamma M_2^2\over M_1M}.
\label{12}
\ee
The CAML process can only amplify (or damp) an already--existing mass
transfer process driven by angular momentum losses $\dot J_{\rm sys}$
(j--driven systems; $\dot J_{\rm sys}$ is the `systemic' rate given by
e.g.\ gravitational radiation or magnetic braking) or a nuclear
expansion rate $\dot R_2 > 0$ (for n--driven systems). 
Defining the evolution time $t_{\rm ev} = |J/\dot J_{\rm sys}|$ or
$(2R_2/\dot R_2)_{M={\rm const.}}$ 
and repeating the algebra of KK95 gives
\be
\dot M_2 = -{M_2\over Dt_{\rm ev}},
\label{13}
\ee
where 
\begin{eqnarray}
\lefteqn{D =} &  \label{14a}   \\
& {1\over 2} \left[ \zeta + 2 + \beta_2 - \frac{2M_2}{M_1} +
\frac{M_2}{M} + (\eta \Gamma - 1) \frac{M_2}{M_1} \left(\frac{M_1}{M}
- \beta_2 \right) \right] . \nonumber 
\end{eqnarray}
(This is a slight generalization of KK95 in the case of
n--driven systems.)
Here $\beta_2 \equiv {\rm d}\ln f_2/{\rm d}\ln (M_2/M_1)$ is the
logarithmic derivative of the ratio $f_2=R_L/a$ (the donor's Roche
lobe radius $R_L$ in units of the binary separation) with respect to
the mass ratio. If $M_2 \ga M_1$ as in KK95 we have 
$\beta_2\simeq - M_1/3M$, hence 
\be
D = {5\over 6} + {\zeta\over 2} - {M_2\over M_1} + \eta \Gamma
{2M_2\over 3M}, 
\label{14}
\ee
which is eq.~(16) of KK95 with $\alpha$, $\nu$ form 
our eq.~(\ref{12}).
In general, the donor's mass--radius index $\zeta = \partial \ln
R_2/\partial \ln M_2$ appearing in 
(\ref{14a},\ref{14}) depends on the secondary's full internal
structure and cannot be expressed in closed form. But in many cases an
estimate  
is available. For slow mass transfer and j--driven or n--driven
evolution with a main--sequence donor (case A) we have $\zeta \simeq
\zeta_{\rm eq} \simeq O(1)$ ($\zeta_{\rm eq}$ is the mass--radius
exponent evaluated under the assumption of thermal equilibrium). 
For n--driven case B evolution slow mass transfer corresponds to 
$\zeta \simeq 0$.
If mass transfer is rapid $\zeta$ approaches $\zeta_{\rm ad}$, the
mass--radius index evaluated with constant entropy profile in the
star. Fully convective stars or stars with a deep convective envelope
have $\zeta_{\rm ad}=-1/3$.
Figure~5 shows $D$ from (\ref{14a}) as a function of mass ratio, for 
various values of $\Gamma$ (with $\zeta=0$, $\eta=1$). Eggleton's
approximation for $f_2$ (Eggleton 1983) was used to calculate
$\beta_2$.

Comparing with the case with no CAML (i.e. $\alpha = \nu = 0$;
conservative mass transfer)  
\be
D=D_0 = {5\over 6} + {\zeta\over 2} - {M_2\over M_1}
\label{15}
\ee
(for $M_2\la M_1$) 
and neglecting any change in $\zeta$ and $t_{\rm ev}$
we see that the mass loss from the black hole always reduces the mass
transfer rate (by trying to widen the binary).  With CAML 
the Roche lobe expands faster (or contracts slower) than without CAML
for a given transfer rate; hence $|\dot M_2|$ adjusts to a smaller
value.     
For $M_2$ significantly smaller than $M_1$ the reduction in $-\dot
M_2$ is small unless 
\be
\eta\Gamma \gg {3M\over 2M_2}. 
\label{16}
\ee 
If this holds we have
\be
\dot M_2 \simeq -{3M\over 2\eta\Gamma t_{\rm ev}} .
\label{17}
\ee

Figure~6 shows the variation of the reduction factor $D(\Gamma)/D_0$
with mass ratio (assuming $\zeta=0=$~const., $\eta=1$). 
The reduction is largest close to where $D \rightarrow 0$ with
conservative mass transfer.
This signals instability against dynamical--timescale 
(if $\zeta=\zeta_{\rm ad}$) or thermal--timescale mass transfer (if
$\zeta=\zeta_{\rm eq}$), with ensuing transfer rates much in excess of
values indicated by (\ref{13}) with $\zeta=O(1)$. 
A very high transfer rate may cause a common envelope phase which
could destroy the system (but see King \& Ritter 1998). 
However, jets actually stabilize mass transfer in systems
where a 
large mass ratio $q=M_2/M_1$ would make mass transfer unstable ($D_0 <
0$) without the jet losses. 
Mass transfer is stable if the mass ratio is smaller than $q_{\rm
crit}$, where $D(q_{\rm crit})=0$. From (\ref{14}) we obtain the
positive root of $D=0$ as   
\be
q_{\rm crit} = {1\over 2}\biggl[C+{2\eta\Gamma\over 3} - 1 + 
\sqrt{(C+{2\eta\Gamma\over 3} -1)^2 + 4C}\biggr],
\lb{16a}
\ee
where $C = 5/6 +\zeta/2$. 
A reasonable approximation for $-1/3\la \zeta\la 10$ is 
$q_{\rm crit} \simeq (\eta \Gamma /2.82)^2 + (5+3\zeta)/6$. 
Systems that are unstable against conservative mass transfer but
stabilized by jet--induced CAML will still encounter 
the instability at the end of the jet phase. 
The mass loss rate from the hole is much larger than the mass
loss rate from the donor, so that the mass ratio $M_2/M_1$ increases
during the jet phase.

The maximum duration $\Delta t$ of the jet--induced CAML phase
depends only weakly on
$\eta \Gamma$, unlike the transfer rate. As shown above, the extractable
gravitating mass from a Kerr black hole with gravitating mass $M_1$ 
is limited by $\Delta m M_1$, where $\Delta m$ depends on the
efficiency of the extraction process.  
The corresponding limit $\Delta M_{\rm tr} < \Delta m
M_1/(\eta\Gamma)$ for the transferred rest--mass during the jet phase
translates into an upper limit $\Delta t < \Delta M_{\rm tr}/(-\dot
M_2)$ for the duration of this phase. 
Using (\ref{17}) as a
representative value for $\dot M$, and 0.1 as a typical value for
$\Delta M$, we have   
\be
\Delta t \la 0.1 \frac{M_1}{M} t_{\rm ev}. 
\label{18}
\ee

As is
well known (van Paradijs 1996; King, Kolb \& Burderi 1996; King, Kolb
\& Szuszkiewicz 1997) a low
transfer rate is a necessary condition for the system to appear as a
soft X--ray transient. This is both empirically true, and expected
from the disc instability picture.
Black--hole systems with low $M_2$ all have 
$-\dot M_2 \sim M_2/t_{\rm ev}$ small enough to satisfy this, whether
j--driven or n--driven (King, Kolb \& Szuszkiewicz, 1997), i.e. all
low--mass black hole systems are transient, even without the extra
effect of the jet losses. Values $\Gamma \gta 10$ could however allow
a black hole system with a higher--mass companion nevertheless to
appear as a transient.

A possible case in point is GRO~J1655-40, whose secondary may have a mass
$M_2\simeq 2\msun$ (Orosz \& Bailyn 1997;  van der Hooft et al.\
1998; Phillips et al.\ 1998) 
and appears to be crossing the Hertzsprung gap. Although the secondary
is close to a regime where its radius expansion temporarily slows,
producing a low transfer rate and transient behaviour (Kolb et al.,
1997, Kolb 1998), current estimates of its parameters  
consistently show it to be too hot to be in this
regime. Instead one expects $t_{\rm ev} \sim t_{\rm KH} \sim 10^7$~yr,
where $t_{\rm KH}$ is the thermal timescale of the secondary's
main--sequence progenitor, so $-\dot M_2 \sim M_2/t_{\rm ev} \sim
{\rm few}\times 10^{-7}/D\  \msun {\rm yr}^{-1}$. Without the CAML
effect of the jet
losses we would have $D\sim 1$, making $-\dot M_2$ far higher than the
estimated critical rate $\dot M_{\rm crit} \simeq 4\times 10^{-8}\
\msun {\rm yr}^{-1}$ (from King, Kolb \& Szuszkiewicz, 1997; eqn. 7).
However, if the jets are very energetic ($\Gamma \ga 50$) the
transfer rate would be reduced by more than the required factor
$\simeq 10$ for the claimed mass ratio $\simeq 0.3$ (Orosz \& Bailyn
1997), see Fig.~5. Hence if $\Gamma \gta 50$ the system would appear
as a transient. This of course requires the jets to be considerably
more relativistic than implied by their bulk motion ($\gamma \simeq
2.55$). The occurrence of transients of this type then
depends purely on the physics of the jets, and cannot be predicted
with current theory.
On the other hand, if the donor mass in GRO~J1655-40 is close to the
lower limit $1.4\msun$ found by Phillips et al.\ (1998) the transfer
rate would be well below the critical rate for transient behaviour
even in the absence of jet--induced CAML. 
However, the luminosity given by the spectral type and orbital
period is too high for a $1.4 \msun$ donor in the phase of
crossing the Hertzsprung gap (e.g.\ Kolb 1998). This suggests that the
actual donor mass is nearer to the upper limit ($2.2\msun$) quoted by
Phillips et al.\ (1998).
(We note that Reg\H os et al.\ (1998) suggested that the donor in
GRO~J1655-40 could be still in the core hydrogen burning phase if the
main sequence is significantly widened by convective overshooting in
the star. This would also give a transfer rate smaller than 
the critical rate for any $\Gamma$). 

The jet source SS433 could be affected by jet--induced CAML as well. 
The nature of the compact star in this system is still unclear (e.g.\
Zwitter \& Calvani 1989; D'Odorico et al.\ 1991), but a black hole
cannot be ruled out. The claimed mass ratio of order 3 is certainly
above the stability limit for conservative mass transfer.
However, the donor might not fill its Roche lobe (Brinkmann et al.\
1989), and the jets seem to be less energetic than in the superluminal
sources GRO~J1655-40 and GRS 1915+105.

\section{CONCLUSIONS}

We may draw the following conclusions from the arguments of this 
paper:

1. The mass of a binary black hole changes by only a relatively small amount
during the mass transfer process, whether through accretion
(fractional increase $\lta M_2/M_1$ for a low--mass system, and much
less for a high--mass system) or rotational energy extraction
(fractional decrease $\lta 10$\%). Currently measured masses are
therefore similar to the formation masses.

2. Too little mass is accreted to spin up a hole from $a_* \simeq 0$ to $a_*
\simeq 1$.

3. Extracting angular momentum from the hole can
in principle reduce $a_*\simeq 1$ to $a_* \simeq 0$, but in practice
none of the known systems shows the effects of injecting the extracted
rotational energy of about $10^{54}$ erg into the local
ISM. In combination with 2. above this suggests that binary black
holes also retain a value of $a_*$ close to that at formation. 

4. Jet ejection powered by the black hole's rotational energy
can have a major effect in stabilizing mass transfer,
particularly in higher--mass systems. Sufficiently energetic jets
can reduce the mass transfer rate to values making the system transient.

If we accept 3. above, it would appear that any claimed range of $a_*$
or $M_1$ must represent the range of initial conditions for binary
black holes. 
However, 
a simple extrapolation of the neutron star case seems to favour 
values $a_* \simeq 1$.
In particular even the maximum angular momentum $J_1 = GM_1c = 3\times
10^{37}(M_1/10\msun)$ cm$^{2}$ s for the hole
is far smaller than any plausible value for that of 
the progenitor star. If confirmed, the range of $a_*$ claimed by Zhang
et al.\ (1997) thus represents a challenge to theory.

\paragraph*{Acknowledgments}
UK would like to thank Ulrich Anzer, Hans Ritter and Maximilian
Ruffert for useful discussions. We thank the referee, Saul Rappaport,
for comments which helped to improve the manuscripts. 
ARK acknowledges support as a PPARC Senior
Fellow. Theoretical astrophysics research at Leicester is supported by
a PPARC rolling grant.

{}

\clearpage

\onecolumn

\begin{table*}
\begin{minipage}{120mm}
\caption{\bf Binaries with claimed measurements of black--hole spin
$a_*$.}
\bigskip

\begin{tabular} {llllll}
\hline
system & orbital period & $a_*$ & type & $M_{\rm tr}/\msun$ & $M_{\rm
spin-up}/\msun$ \\ 
\hline
GS$1124-68$    &  10.4 h & -0.04  & j--driven & $< 1.5$  & -     \\
GS$2000+25$    &  8.3 h  &  0.03  & j--driven & $< 1.5$  & -     \\
LMC X-3      &  1.7 d  & -0.03  & n--driven & $\la2.5$   & -     \\
             &         &     & wind--fed    & $\la0.1$   &   \\
GRO~J1655-40 &  2.62 d  & 0.93   & n--driven & $<1.9$  & 3.3   \\
             &         & $>0.7$ &           &          & 1.9   \\
GRS 1915+105 &   ?     & 0.988? & n--driven & $\la5$   & 22   \\
             &         &        & wind--fed & $\la0.1$ &       \\
\hline
\end{tabular}
\vspace{2cm}
~
\end{minipage}
\end{table*}


\vspace*{5cm}

\section*{Figure Captions}

{\bf Figure~1} {\em Top panel:} Estimated upper limit $M_{\rm acc}$
for the mass a 
black hole can accrete during case A mass transfer, as a
function of initial donor mass $M_{2,{\rm initial}}$ (assuming that
the donor is initially on the ZAMS and the black hole has mass $8\msun$). 
{\em Middle panel:} Final period $P_{\rm f}$ (solid, scale on
the left) and initial period $P_{\rm i}$ (dashed, scale on the right).
{\em Bottom panel:} Total duration $t$ of mass transfer phase
(ending when the donor reaches the terminal main sequence; solid, scale
on the left) and final secondary mass $M_{2,{\rm f}}$ (dashed,
scale on the right).  
The donor's mass--radius index $\zeta$ was fixed at 0.  
\bigskip

{\bf Figure~2} As Fig.~1, but for case B mass transfer, with
the donor initially on the terminal main-sequence. Mass transfer
terminates when core helium burning begins.

\bigskip

{\bf Figure~3}
Black hole spin $a_*$ vs.\ accreted rest--mass $\Delta M$, in units
the final mass $M_{\rm max}$ at maximum spin (solid line, bottom axis),
and in units of the initial mass $M(a_*=0)$ at zero spin 
(dashed, top axis).

\bigskip

{\bf Figure~4}
Extracted fractional energy $\Delta m=(M_{\rm max} - M_1)/M_{\rm max}$
as a function of efficiency $\epsilon$ of extraction of a black hole's
rotational energy, cf.\ (\ref{10a}).

\bigskip

{\bf Figure~5}
Stability term $D$ from eq.~({\ref{14a}) vs.\ mass ratio $q=M_2/M_1$,
for jets with different specific energy $\Gamma$ (assuming $\zeta=0$).

\bigskip

{\bf Figure~6}
Mass transfer reduction factor $D(\Gamma)/D_0$ vs.\ mass ratio
$q=M_2/M_1$, for jets with different specific energy $\Gamma$
(assuming $\zeta=0$). 

\newpage

\begin{figure*}
\plotone{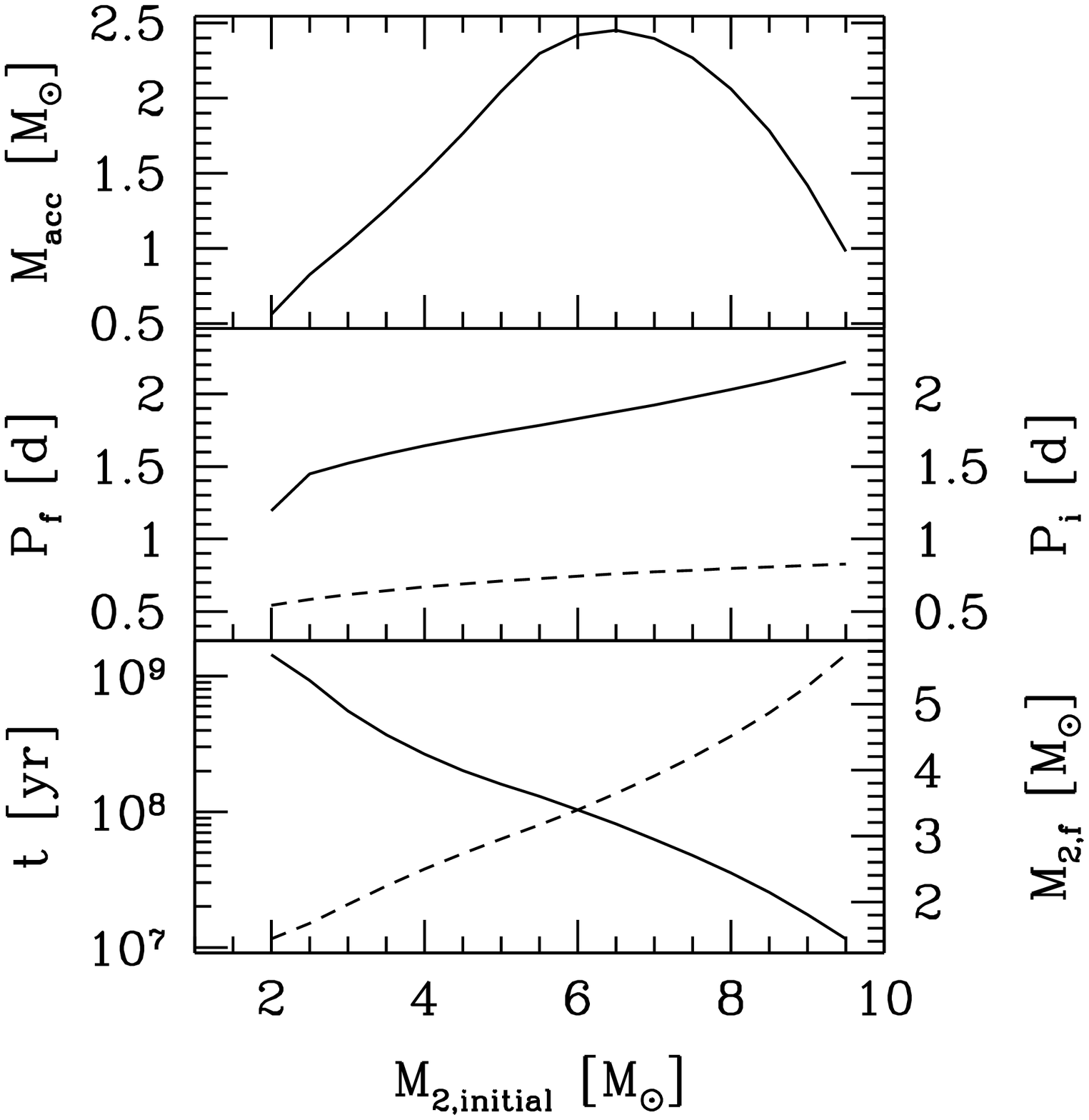}

\vspace{1.5cm}

{\huge Figure 1}
\end{figure*}

\begin{figure*}
\plotone{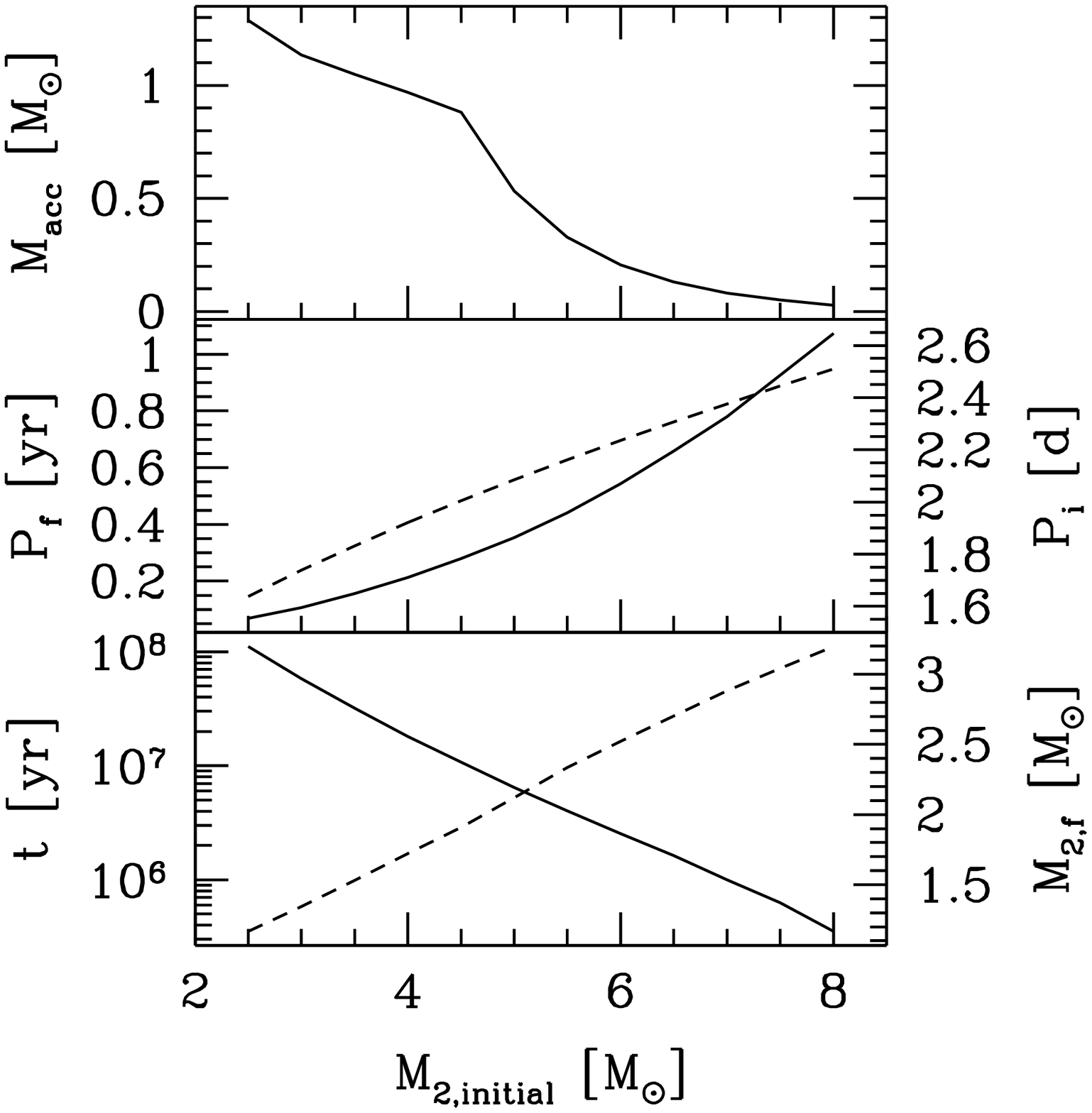}

\vspace{1.5cm}

{\huge Figure 2}
\end{figure*}

\begin{figure*}
\plotone{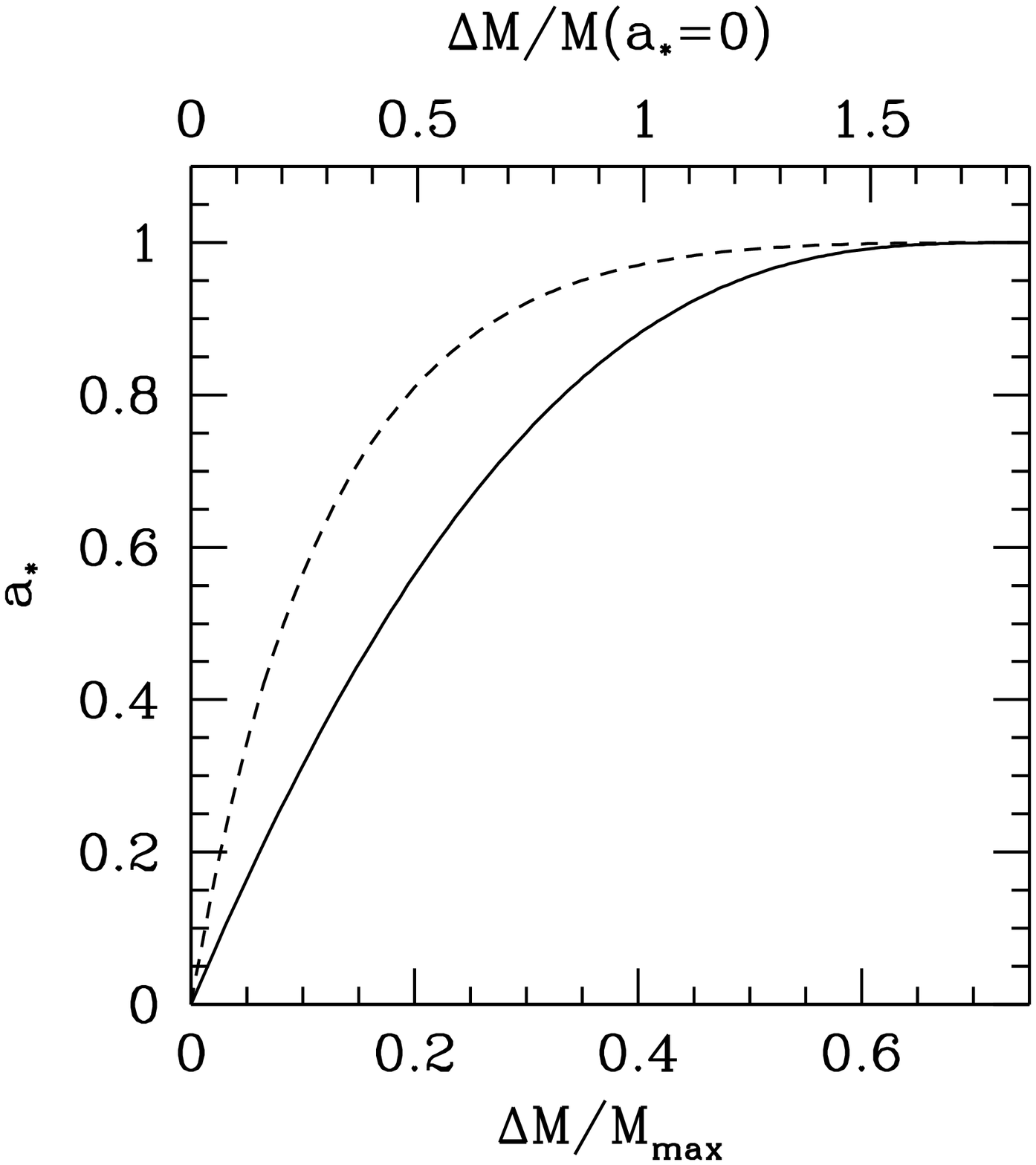}

\vspace{1.5cm}

{\huge Figure 3}
\end{figure*}

\begin{figure*}
\plotone{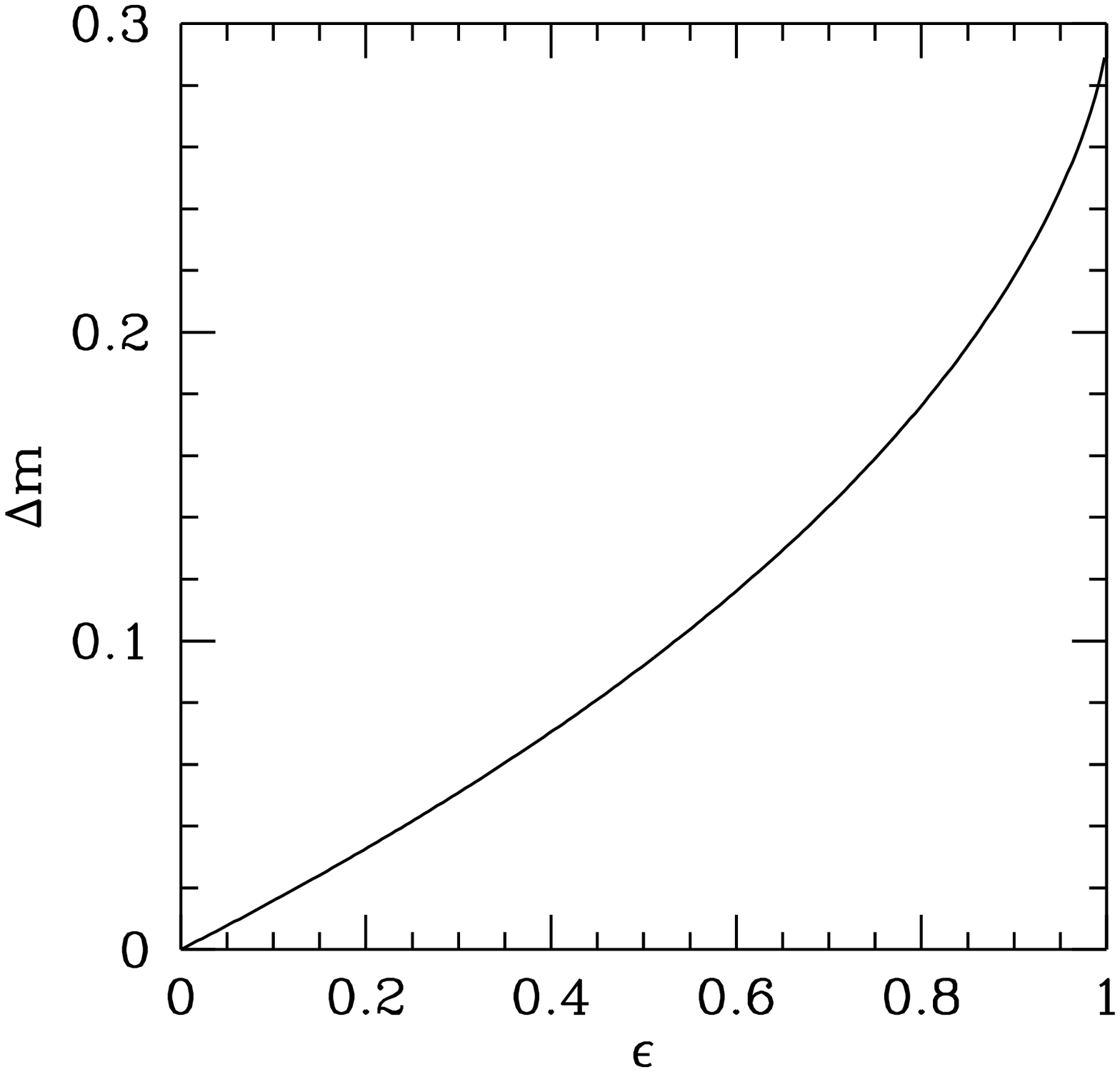}

\vspace{1.5cm}

{\huge Figure 4}
\end{figure*}

\begin{figure*}
\plotone{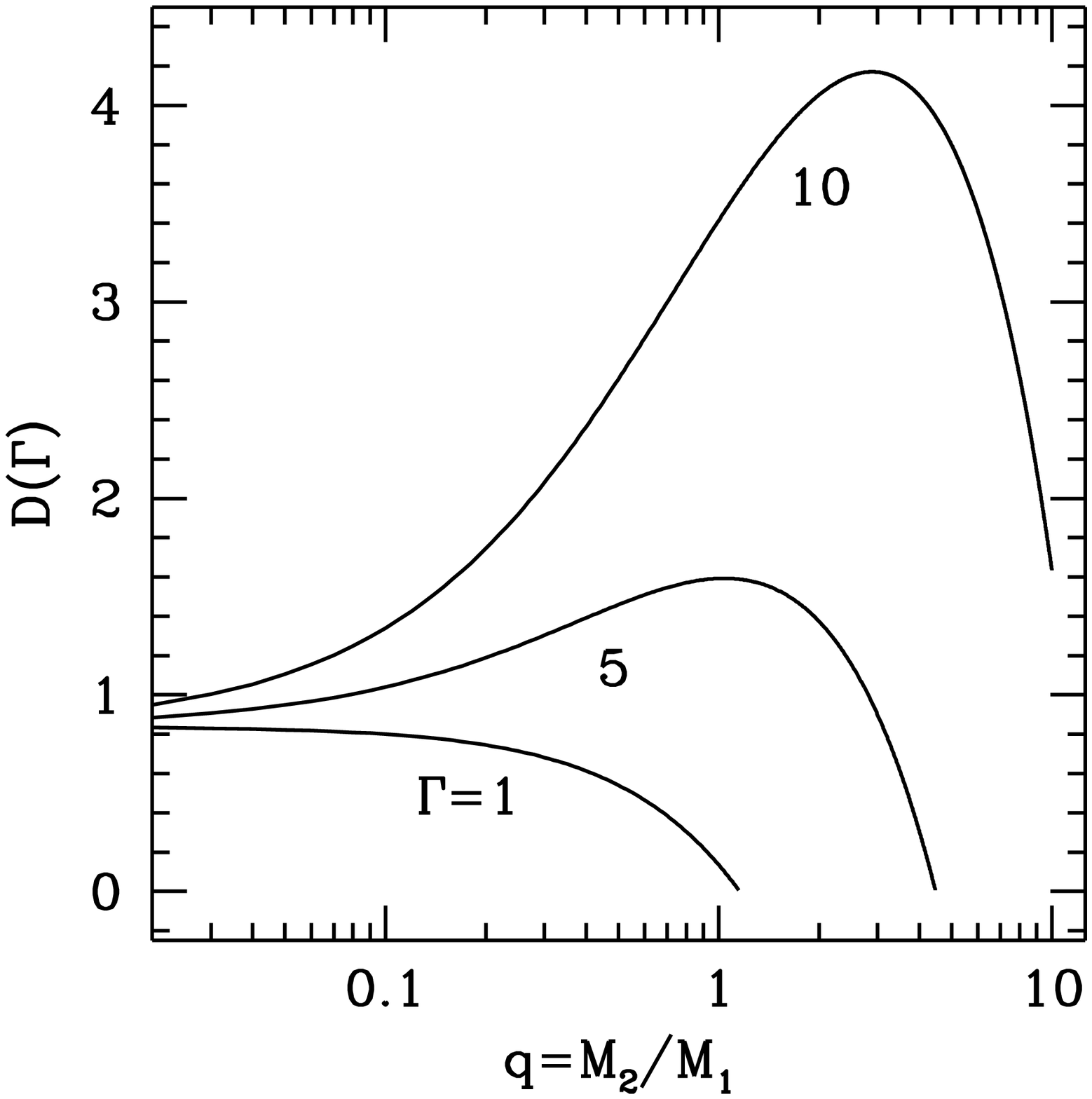}

\vspace{1.5cm}

{\huge Figure 5}
\end{figure*}

\begin{figure*}
\plotone{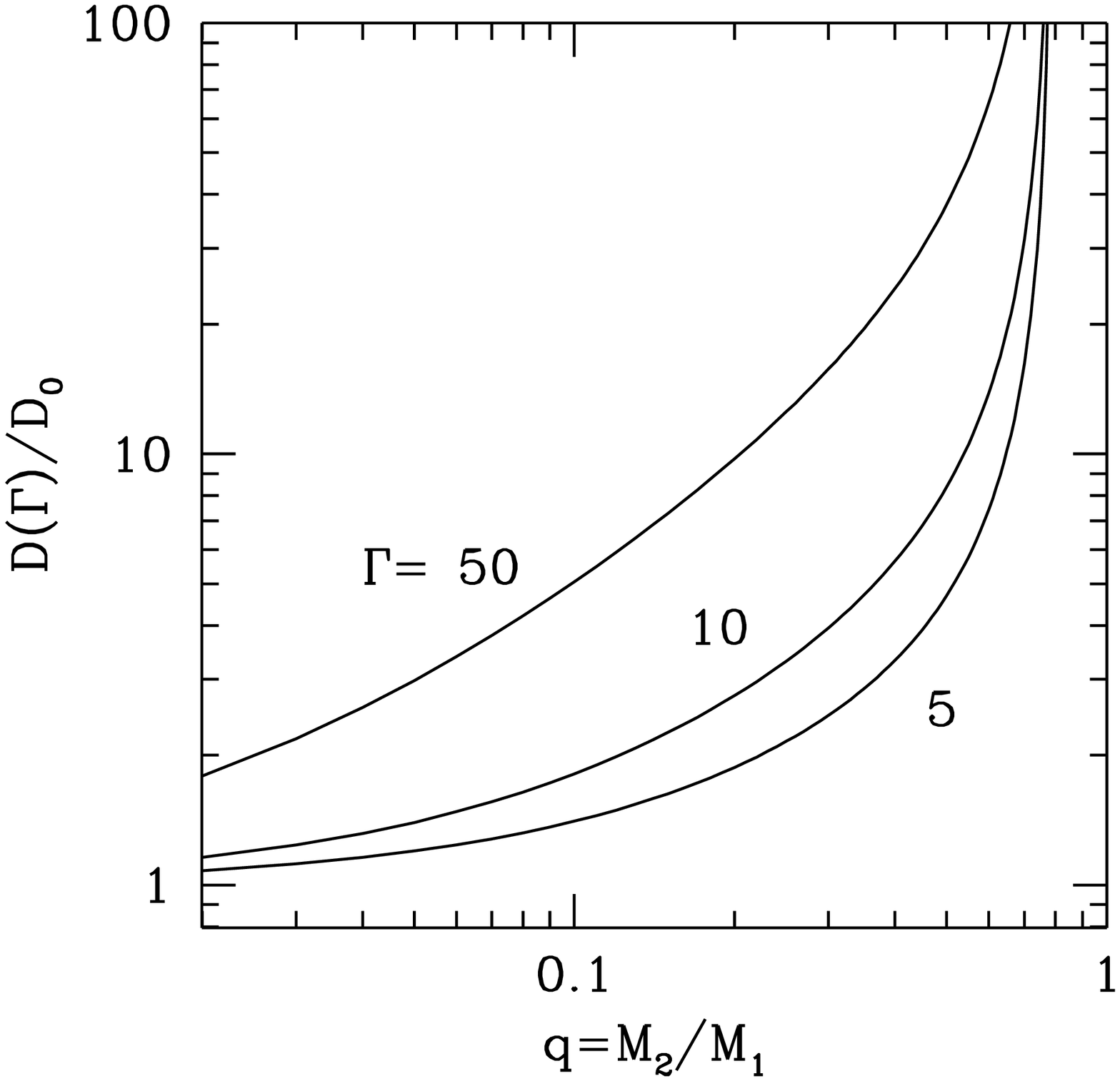}

\vspace{1.5cm}

{\huge Figure 6}
\end{figure*}

\end{document}